\documentclass[11pt,a4paper]{article}
\usepackage[utf8]{inputenc}
\usepackage[T1]{fontenc}
\usepackage{lmodern}
\usepackage{pdfpages}
\usepackage{amsmath,amssymb,graphicx,booktabs,tabularx}
\usepackage[left=22mm,right=22mm,top=22mm,bottom=22mm]{geometry}
\usepackage[numbers,sort&compress]{natbib}
\usepackage{microtype,xcolor,caption,placeins}
\usepackage{setspace}
\usepackage{ragged2e}
\usepackage[skip=0.65\baselineskip]{parskip}
\usepackage{hyperref}
\input{si_labels}
\hypersetup{colorlinks=true,linkcolor=black,citecolor=black,urlcolor=blue}
\newcommand{\wt}{\,wt.\%}

\title{Prediction targets and surrogate models for active learning of spinel oxygen carriers}
\author{Alfred Juhlin Onbeck$^{1}$, Ivana Staničić$^{1}$, Radhey Shyam Yadav$^{1}$,\\ 
Tobias Mattisson$^{1}$, Anders Hellman$^{2,3,*}$}
\date{
    \small{$^{1}$Department of Environmental and Energy Sciences, Division of Energy Technology, Chalmers University of Technology, Chalmersplatsen 4, Gothenburg, SE-41296, Sweden} \\
    \small{$^{2}$Department of Physics and Astronomy, Division of Chemical Physics, Chalmers University of Technology, Chalmersplatsen 4, Gothenburg, SE-41296, Sweden} \\[0.5em]
    \small{$^{3}$Competence Centre for Catalysis, Chalmers University of Technology, Chalmersplatsen 4, Gothenburg, SE-41296, Sweden} \\[0.5em]
    \small{$^{*}$Correspondence: anders.hellman@chalmers.se} 
}

\begin{document}
\RaggedRight
\maketitle

\begin{abstract}
Active learning accelerates materials discovery by using a surrogate model to choose which candidates to evaluate next. Comparisons usually focus on the type of surrogate model, with less attention to what the model is trained to predict. The target property is often a composite function of expensive physical quantities that a surrogate could predict instead. Here we compare the choice of surrogate with the choice of prediction target in a search for multicomponent spinel oxygen carriers for chemical looping, where the oxygen transfer capacity (OTC) is calculated from Gibbs energies across oxygen vacancy counts. Ten strategies, each run five times, pair one surrogate (a deep ensemble, a Gaussian process or Monte Carlo dropout) with one target: OTC itself, OTC capped at a composition's maximum capacity, or, for the first two surrogates, the underlying Gibbs energies. What the surrogate predicted mattered far more than which surrogate was used. With OTC as the target, all three surrogates performed nearly identically, whereas predicting the Gibbs energies raised the mean capacity acquired per run by 0.65--0.81 percentage points and produced 279 of the 281 acquisitions at or above 6\,wt.\%. Capping direct OTC predictions improved average performance, particularly for the Gaussian process, but still rarely found the highest-capacity materials. Because maximizing OTC rewards low molar mass, every search moved toward light, Li- and Al-rich compositions, whose highest-capacity members mostly need unusually high oxidation states. For targets derived from intermediate physical quantities, predicting those quantities may matter more than the choice of surrogate.
\end{abstract}

\section{Introduction}

    Materials discovery has become increasingly data-driven, where identified structure–property relationships are used to guide the search for new candidates~\cite{Butler2018, Norskov2009, Yao2022}.
    This approach makes it possible to systematically screen compositional and structural spaces beyond the usual trial-and-error approach, and as such has shown promising results in different materials technologies, for example, in catalysis and other energy-related applications~\cite{Studt2014, GomezBombarelli2018, Hus2019, Zhong2020, Zeni2025}.
    The introduction of machine learning has further accelerated this process, enabling the rapid screening of millions of candidates~\cite{MeBaSc2023}.
    As a result, materials discovery is moving toward automated pipelines~\cite{Tabor2018, PyzerKnapp2022}, in which the generation of material candidates and the systematic evaluation of the desired property, whether through first-principles calculations, machine-learning interatomic potentials, or experiments, play central roles.

    However, when property evaluation is too expensive, active learning (AL)~\cite{Xue2016, Lookman2019, Rao2022, Moon2024, RaBaZh2024} offers an efficient alternative to exhaustive search.
    The central component of active learning is a surrogate model trained on an initial dataset, which acts as a fast approximation of the expensive evaluation, allowing many candidates to be screened at a fraction of the cost of, for example, density functional theory (DFT) calculations or experiments.
    The surrogate is then used to guide which compositions to evaluate next, balancing exploitation of promising known regions with exploration of uncertain areas.
    Although AL has not been explored as comprehensively in the materials informatics literature as supervised learning on fixed datasets, it has proven especially useful for navigating complex compositional spaces with limited training data, including high-entropy alloys and multi-metal oxides~\cite{Rao2022, BRORSSON2026102239, Moon2024}.

    AL performance depends on the choice of surrogate model~\cite{chen2026}, but also on what that model is trained to predict.
    Many properties targeted in materials discovery are calculated from intermediate physical quantities, for example catalyst performance from adsorption energies~\cite{Tran_2018} and thermodynamic stability from the formation energies of competing phases~\cite{MeBaSc2023, Novick_2024}.
    For such properties, the surrogate can either predict the property directly or predict the intermediate quantities and calculate the target property from them afterwards.
    In Bayesian optimization of composite functions, learning the intermediate quantities in this way can improve sampling efficiency~\cite{Astudillo2019}, and the approach has also been applied to material and structural design~\cite{Cardoso_Coelho_2025}.
    When the calculation from the intermediate quantities also enforces a known upper limit for each candidate, predictions derived from those quantities automatically stay within that ceiling, whereas direct predictions of the target property can exceed it.
    In each cycle, candidates are ranked for evaluation by an acquisition score, which here is a target value drawn from the surrogate's predictive distribution (or calculated from a drawn set of intermediate quantities).
    To test whether predicting the intermediate quantities helps mainly by respecting the upper limit or by learning the underlying quantities themselves, we also test clipping the direct acquisition score at each candidate's ceiling before selection.
    We examine how these two choices, the prediction target and acquisition-score clipping, affect oxygen-carrier discovery across different surrogate models.

    As a case study, we consider spinel oxygen carriers for chemical-looping combustion, a combustion process with inherent CO$_2$ capture~\cite{adanez2012progress, Lyngfelt2020}.
    A solid oxygen carrier circulates between two reactors. In the fuel reactor (FR), it gives up lattice oxygen to oxidize the fuel, and in the air reactor (AR), it is reoxidized by air.
    Because the fuel never mixes with air, the CO$_2$ leaves the FR without nitrogen dilution.
    The property we optimize is the carrier's oxygen transfer capacity (OTC), the mass fraction of oxygen it releases per cycle, set by the number of oxygen vacancies that open up between the AR and FR conditions (Section~\ref{sec:otc}).
    A high OTC is desirable because less carrier then has to circulate between the reactors to deliver the oxygen the fuel needs.
    For comparison, the redox couples used in practice have stoichiometric capacities of 3.4\,wt.\% for Fe$_2$O$_3$/Fe$_3$O$_4$, 7.0\,wt.\% for Mn$_3$O$_4$/MnO and 10.1\,wt.\% for CuO/Cu$_2$O~\cite{adanez2012progress}.
    Practical carriers transfer less oxygen per unit mass because synthetic carriers usually disperse the active oxide on an inert support, and natural ores contain other minerals~\cite{adanez2012progress, Lyngfelt2020, Daneshmand_Jahromi_2023}.
    Oxygen carriers have progressed toward more complex metal oxides over the past decade, and high-entropy oxides have only recently been explored~\cite{ADANEZRUBIO2024131307, shen2025high, BRORSSON2026102239, zhao2026entropy, li2026chemical, hu2026high, cai2026nicozn}.
    Systematic investigation is difficult because, even with a single crystal structure, a limited set of cations and a fixed 84-atom supercell, the number of possible compositions is far too large to evaluate exhaustively.

    This work compares ten active learning strategies built around deep ensembles~\cite{lakshminarayanan2017simple} (DE), Gaussian processes~\cite{Rasmussen2006} (GP) and Monte Carlo dropout~\cite{Gal2016} (MC).
    The strategies either predict OTC directly (referred to as scalar targets), with or without clipping the acquisition score at the composition-dependent capacity ceiling, the OTC a composition reaches when it releases all seven modeled vacancies, or predict the underlying Gibbs energies (referred to as Gibbs targets), either as a vector over vacancy counts or as a function of composition and vacancy count.
    DE and GP are tested with all four targets and MC with the two scalar ones, and each strategy is run five times with the same budget of 200 selections per run.
    We also compare them with random selection and with a search that ranks candidates by the capacity ceiling alone, and test how well the DE and GP surrogates predict the OTC of a common set of compositions not used in their training.
    We measure how the prediction target and clipping affect the average quality of an acquisition, the number of high-OTC discoveries and the compositions the search moves toward.
    The main result of this study is that the prediction target affects discovery performance far more than the choice of surrogate model, as every run with a Gibbs target acquired a higher mean capacity than every unclipped scalar run, and nearly all discoveries at or above 6\,wt.\% came from the Gibbs targets.
    Clipping the score of the direct OTC prediction narrows the gap to the Gibbs targets, closing most of it for GP but only about a third for DE, and clipped runs rarely reach the highest capacities.

\section{Methods}

    \subsection{Active learning protocol}
    \label{sec:al_protocol}

\begin{figure}[htbp]
            \centering
            \includegraphics[width=\textwidth]{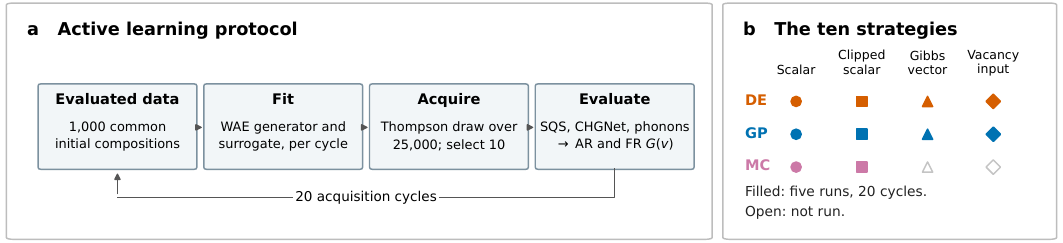}
            \caption{Active learning protocol and tested strategies. (a)~Active learning protocol of one run. Starting from the 1,000 common initial compositions, each of the 20 acquisition cycles refits the WAE generator and the surrogate, scores 25,000 generated candidates with one Thompson draw each, and evaluates the ten highest-scoring ones. (b)~The ten tested strategies. Each filled marker has five runs, and open markers indicate untested combinations.}
            \label{fig:design}
        \end{figure}

        The active learning loop used here has five components: (i) a database of evaluated compositions, (ii) a surrogate model that predicts the target property and its uncertainty, (iii) a candidate generator that proposes new compositions from the search space, here a Wasserstein autoencoder (WAE, Section~\ref{sec:wae}), (iv) an acquisition function that uses the surrogate's predicted distribution to score these candidates, and (v) an evaluation pipeline that computes the target for the selected candidates and adds them to the database.
        In each cycle, the surrogate and the acquisition function decide which candidates are evaluated next, and the surrogate is retrained on the expanded database.
        We call one complete search by one strategy a run. A run starts from an initial dataset of 1,000 evaluated compositions, shared by all runs. In each of 20 acquisition cycles, it selects ten candidates, evaluates them and adds them to its own database, on which its surrogate and candidate generator are then refitted (Fig.~\ref{fig:design}a), giving a budget of 200 selections per run.
        All strategies in this work use the same acquisition function, Thompson sampling~\cite{Thompson_1933}, which is one of several common choices alongside the upper confidence bound and expected improvement~\cite{Shahriari_2016}.
        It draws one value of the target from the surrogate for each of the 25,000 generated candidates, converts it to an OTC value when the target is a set of Gibbs energies, uses the resulting OTC as the candidate's acquisition score, and selects the ten candidates with the highest scores.
        A candidate with an uncertain prediction is occasionally selected through a high draw, while one with a consistently high prediction is selected reliably, so the rule balances exploration and exploitation without an additional trade-off parameter.

\begin{figure}[htbp]
            \centering
            \includegraphics[width=\textwidth]{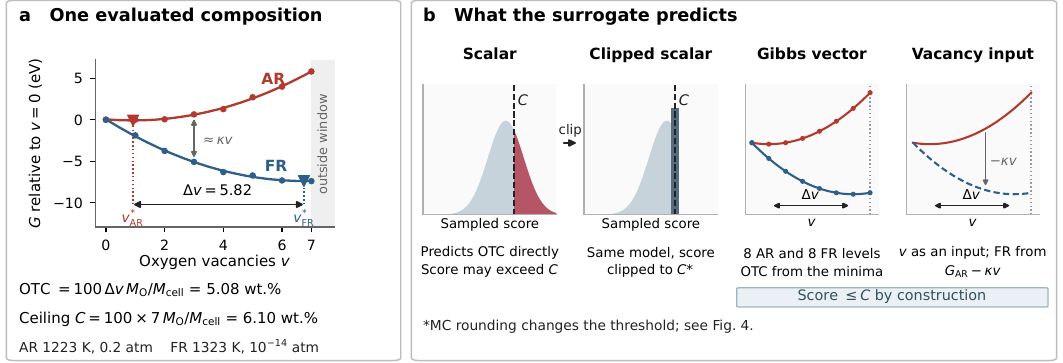}
            \caption{Target definition. (a)~Evaluated Gibbs energies relative to each reactor's vacancy-free reference and quadratic fits for Al$_{14}$CuLi$_{11}$Mn$_{10}$O$_{48}$, acquired in cycle 6 of Gibbs-vector GP run 5. The fitted minima, $v^*_{\mathrm{AR}}=0.924$ and $v^*_{\mathrm{FR}}=6.746$, give OTC $=5.076$\,wt.\%, below the composition's ceiling of $6.103$\,wt.\%. (b)~Schematic target representations. The ceiling $C(x)$ is the OTC a composition would reach by releasing all seven modeled vacancies (equation~\eqref{eq:capacity_ceiling}). Gibbs-target strategies stay within this limit because their vacancy counts lie in the modeled interval, and scalar clipping applies it to the sampled score. The scalar distributions are illustrative.}
            \label{fig:target}
        \end{figure}

        Each strategy pairs a surrogate model with one of four targets (Fig.~\ref{fig:target}b). The target defines what the surrogate is trained to predict and how its sampled prediction is turned into the OTC score by which candidates are ranked, so all strategies search for high OTC but obtain the score in different ways:
        \begin{itemize}
            \item \textbf{Scalar:} predict OTC directly.
            \item \textbf{Clipped scalar:} predict OTC directly and cap the sampled score at the composition's capacity ceiling (Section~\ref{sec:clipping}).
            \item \textbf{Gibbs vector:} predict all 16 Gibbs energies, at the eight vacancy counts $v=0$--$7$ in both reactors, together as one multi-output prediction.
            \item \textbf{Vacancy input:} predict the AR Gibbs energy from the composition and the vacancy count, evaluate it at $v=0$--$7$, and obtain the FR energies by subtracting, for each vacancy, the change in oxygen chemical potential between the reactors (Section~\ref{sec:vacancy_input}).
        \end{itemize}
        The Gibbs-vector and vacancy-input targets both predict intermediate Gibbs energies rather than OTC directly, and we refer to strategies that use them as Gibbs-target strategies. Both convert the predicted energies for each reactor to an equilibrium vacancy count at the minimum of a quadratic fit over $v\in[0,7]$ (Fig.~\ref{fig:target}a). Because these vacancy counts lie within the modeled interval, the resulting OTC scores cannot exceed the capacity ceiling, so clipping is applied only to the scalar target.
        Pairing the three surrogates with these four targets gives ten strategies (Fig.~\ref{fig:design}b): DE and GP are tested with all four targets, whereas MC is tested only with the two scalar ones. In MC, the surrogate is the regression decoder of the WAE that also generates the candidates (Section~\ref{sec:mc}), so replacing the scalar target with Gibbs energies would also change the candidate generator and confound the two effects.

        The shared initial dataset of 1,000 compositions, denoted cycle 0, was generated together with the 500 additional compositions of the random reference. For each composition, four to seven cation elements were drawn, with at least one on each sublattice, and the 12 A-site and 24 B-site cation counts were drawn at random from the chosen elements on each sublattice. Compositions that passed the tolerance-factor and charge-neutrality screen were then built as special quasi-random structures (SQS), disordered cation arrangements with minimal short-range order generated with icet~\cite{icet}, in the same way as the candidates acquired during the search.

        Each strategy is run five times. Every DE and GP run adds 200 new evaluated compositions to its database, whereas the five MC runs add 995 in total (198 to 200 per run) and the five clipped-MC runs add 987 (197 to 198 per run), giving 9,982 new evaluation records across the comparison. The 18 selections that did not yield a new evaluation are itemized in SI Section~\ref{SI-si:screening}.

        The five runs of each strategy use random seeds 1 to 5, so they draw different candidates. At a given seed, the clipped and unclipped scalar strategies of the same surrogate fit their WAE on identical data and use the same surrogate in the first cycle, so they draw from the same candidate pool and partly select the same candidates (SI Section~\ref{SI-si:configs}). At the same seed, no other two strategies share any first-cycle selection, and their candidate pools can differ because the OTC that the WAE trains on is scaled differently between groups of strategies (SI Section~\ref{SI-si:configs}). Comparisons between a clipped strategy and its unclipped counterpart therefore match runs by seed, whereas all other comparisons treat the runs as independent (Section~\ref{sec:statistics}).

    \subsection{Spinel composition space}
    \label{sec:composition_space}

        The compositions have the spinel structure AB$_2$O$_4$, with A-site cations on tetrahedral and B-site cations on octahedral sites. The A-site cations are restricted to Li, Mg, Zn, Cu, Co, Ni and Ca, and the B-site cations to Al, Cr, Ga, Ti, Fe, Mn, Zr, Sn, In and V.
        Most of these elements were selected because they have been incorporated in synthesized high-entropy spinels, including Mg, Zn, Cu, Co and Ni~\cite{musico2019tunable, fracchia2020eight, MANDAL2025177993, corlett2023synthesis}, Li~\cite{duan2021new, ito2026suppression}, Al, Cr, Fe and Mn~\cite{dabrowa2018synthesis, musico2019tunable, fracchia2020eight, mao2023high}, Ga~\cite{johnstone2022entropy, li2026gallate}, Ti~\cite{ganjali2025electrochemical}, V~\cite{ito2026suppression} and Sn~\cite{kaplan2025tuning}. Ca and Zr have been reported in only a few high-entropy spinels~\cite{qiao2023effect, li2026chemical}, and In has only been reported in simpler spinels, so including them expands the search space beyond the high-entropy spinels already tested experimentally.
        Each SQS is built in an 84-atom supercell formed by $1\times2\times3$ repeats of the 14-atom primitive cubic spinel cell ($Fd\bar{3}m$, Materials Project~\cite{jain2013} prototype MgAl$_2$O$_4$, mp-3536), containing 12 A sites, 24 B sites and 48 oxygen sites. The vacancy count $v$ ranges from 0 to 7, so that $v/48$ is the fraction of vacant oxygen sites.

        Before screening, candidates with more than eight cation species and those that duplicate the database or another candidate are removed. The remaining candidates are screened in descending order of acquisition score until ten pass. Each is mapped to 12 A-site and 24 B-site cations and passes when its spinel tolerance factor~\cite{spinel_tol_factor}
        \begin{equation}
            \tau=\frac{\sqrt{3}}{2}\frac{\bar r_B+\bar r_{\mathrm O}}{\bar r_A+\bar r_{\mathrm O}},
        \end{equation}
        computed from the average ionic radii of each sublattice, satisfies $0.77\leq\tau\leq1.06$, a range chosen to cover the known oxide spinels surveyed in the same work, and its assigned formal charges sum to zero.
        Charge neutrality is checked by integer linear programming, which searches for an assignment of oxidation states to the cations whose formal charges sum to zero. Each cation can take any positive state with a tabulated Shannon radius~\cite{shannon1976} (SI Table~\ref{SI-tab:screen_states}). This set is much wider than the states usual in spinels and includes, for example, Mn(VII) and Cr(VI). When several neutral assignments exist, fixed tie-breaking rules (SI Section~\ref{SI-si:screening}) select one, and the radii of the selected states enter $\tau$.
        The tolerance factor is a coarse geometric proxy for spinel stability, so together the two checks remove compositions whose cation sizes or charges do not fit the spinel structure, although they do not test stability against competing phases.

        For the composition analysis, we report the molar cation mixing entropy per AB$_2$O$_4$ formula unit,
        \begin{equation}
            \frac{S_{\mathrm{mix}}}{R}=-\sum_i x_{A,i}\ln x_{A,i}-2\sum_i x_{B,i}\ln x_{B,i},
            \label{eq:composition_entropy}
        \end{equation}
        where $R$ is the molar gas constant and $x_{A,i}=n_{A,i}/12$ and $x_{B,i}=n_{B,i}/24$ are element fractions on the two cation sublattices.
        This descriptor distinguishes elements but not their oxidation states, and it differs from the configurational entropy $S_{\mathrm{config}}(v)$ in the Gibbs energy, which also includes oxidation states and oxygen/vacancy mixing.

    \subsection{Target property: oxygen transfer capacity}
    \label{sec:otc}

        The OTC quantifies the mass fraction of lattice oxygen that can be released between the AR and FR conditions:
        \begin{equation}
            \text{OTC} = 100\frac{M_{\text{O}} (v_{\text{FR}}^* - v_{\text{AR}}^*)}{M_{\text{spinel}}},
        \end{equation}
        where $M_{\text{O}}=15.999$~g/mol, $M_{\text{spinel}}$ is the molar mass of the 84-atom supercell without vacancies ($v=0$), and $v_{\text{AR}}^*$ and $v_{\text{FR}}^*$ are the vacancy counts that minimize the Gibbs energy of the supercell at the AR and FR conditions (Sections~\ref{sec:gibbs_pipeline} and~\ref{sec:quadratic_fit}).
        At fixed temperature and oxygen partial pressure, the equilibrium oxygen content of the carrier is the one that minimizes its Gibbs energy, including the chemical potential of the released oxygen, so $v_{\text{AR}}^*$ and $v_{\text{FR}}^*$ give its oxygen content in each reactor, and their difference the oxygen released per cycle.
        OTC is expressed in wt.\%, and a higher value means more oxygen delivered per unit mass of carrier.
        When a composition releases all seven modeled vacancies ($v_{\text{AR}}^*=0$ and $v_{\text{FR}}^*=7$), its OTC reaches the composition-dependent capacity ceiling
        \begin{equation}
            C(x)=100\frac{7M_{\mathrm O}}{M_{\mathrm{spinel}}(x)}.
            \label{eq:capacity_ceiling}
        \end{equation}
        The seven-vacancy limit, 14.6\% of the oxygen sites ($\delta\approx0.58$ per AB$_2$O$_4$), lies between the largest oxygen deficiency that LiMn$_2$O$_4$ tolerates before it decomposes, $\delta\approx0.2$~\cite{Sugiyama1996}, and the deficiency at which LiMn$_{1.5}$Ni$_{0.5}$O$_4$ transforms to rock salt, $\delta\approx0.65$~\cite{Pasero2008}, while a synthesized high-entropy oxygen carrier has reversibly released 5.5\wt{} of lattice oxygen, about 18\% of its oxygen, without losing its main cubic phase~\cite{ADANEZRUBIO2024131307}. Compositions that transform to another phase before releasing seven oxygen atoms would release less than modeled, whereas those that remain stable beyond seven vacancies could release more, which the single-phase SQS model cannot capture.

    \subsection{Gibbs energy calculation pipeline}
    \label{sec:gibbs_pipeline}

        Each SQS structure is relaxed with CHGNet~\cite{chgnet}, and its vibrational free energy $F_{\mathrm{vib}}$ is obtained from finite-displacement phonon calculations with Phonopy~\cite{phonopy} using CHGNet forces. Every evaluation uses one structure for each vacancy count. The Gibbs energy of a supercell with $v$ vacancies at reactor conditions is
        \begin{equation}
            G(v;T,p_{\mathrm O_2})=E_{\mathrm{CHGNet}}(v)+F_{\mathrm{vib}}(v,T)+E_{\mathrm{corr}}(v)-T S_{\mathrm{config}}(v)+v\mu_{\mathrm O}(T,p_{\mathrm O_2}),
            \label{eq:helmholtz_revision}
        \end{equation}
        in eV per supercell. The correction $E_{\mathrm{corr}}=E_{\mathrm{form}}-E_{\mathrm{CHGNet}}$ converts the static energy to a formation energy referenced to Materials Project data~\cite{jain2013}. It is linear in $v$ with a composition-independent slope of $-4.948$~eV per vacancy, the Materials Project energy of elemental oxygen per atom, and therefore acts as the reference energy of each removed oxygen atom (SI Section~\ref{SI-si:oracle}).
        The configurational entropy $S_{\mathrm{config}}$ counts the mixing of element--oxidation-state populations on each cation sublattice and of oxygen and vacancies on the 48 anion sites (SI Section~\ref{SI-si:oracle}).
        The oxygen chemical potential per atom, $\mu_{\mathrm O}(T,p_{\mathrm O_2})=\mu_{\mathrm O}^{\mathrm{ref}}(T)+\tfrac12 k_{\mathrm B}T\ln(p_{\mathrm O_2}/p^{\mathrm{ref}})$, takes its reference term from NIST--JANAF data~\cite{NIST_JANAF} following Reuter and Scheffler~\cite{reuter2001}.
        The AR uses $T=1223.15$~K and $p_{\mathrm O_2}=0.2$~atm, and the FR uses $T=1323.15$~K and $p_{\mathrm O_2}=10^{-14}$~atm, the conditions used in earlier active learning work on high-entropy perovskite oxygen carriers~\cite{brorsson2026data}.
        The numerical settings are listed in SI Table~\ref{SI-tab:pipeline}, and further details of the pipeline are given in SI Section~\ref{SI-si:oracle}.

        The stored targets are relative to each reactor's vacancy-free value, $G_r(v)-G_r(0)$, which gives 16 Gibbs energies per composition, at eight vacancy counts in each reactor. The OTC is extracted from the two curves by the quadratic fit described next.

    \subsection{Quadratic fitting of the energy curves}%
    \label{sec:quadratic_fit}

        The SQS and the vacancy arrangement are single realizations of a disordered structure, so the computed energies of a composition depend on which arrangement of cations and vacancies is drawn, more so at high vacancy counts, where the number of possible vacancy arrangements, $\binom{48}{v}$, is largest. Each evaluation builds a new structure, so a composition selected in several runs is evaluated with a different arrangement each time. For the 873 compositions selected in more than one run, the median standard deviation of $G_{\mathrm{AR}}(7)-G_{\mathrm{AR}}(0)$ between their evaluations is 0.41~eV per supercell.
        Such variation could shift the discrete minimum of the eight energies by whole vacancies, particularly for flat Gibbs-energy profiles. The minimizing vacancy counts $v^*_{\mathrm{AR}}$ and $v^*_{\mathrm{FR}}$ are therefore taken from a least-squares quadratic fit to the eight energies of each reactor, restricted to the modeled interval $[0,7]$. The fit reduces the spread of OTC between repeated evaluations of the same composition, as shown in SI Section~\ref{SI-si:quadfit}.

    \subsection{Latent representation and candidate generation}
    \label{sec:wae}

        A WAE~\cite{tolstikhin2019wassersteinautoencoders} fills two roles in the active learning loop.
        First, it provides the candidate pool for Thompson sampling: at each cycle, 25,000 latent vectors are decoded into compositions for the run's surrogate to score.
        Second, its regression decoder, which maps a latent vector to an OTC prediction, serves as the surrogate for the MC strategies.
        All strategies use the same WAE architecture, but each run trains its own WAE, which is retrained in every cycle on the run's growing database, starting from the weights of the previous cycle.

        The WAE encodes the A-site (7) and B-site (10) atomic fractions together with the OTC into an 8-dimensional latent space, so the latent representation depends on both composition and the target property.
        A composition decoder and a regression decoder share this latent space.
        At each cycle, a Gaussian distribution is fitted to the encoded training set, and the candidate latent vectors are sampled from it rather than from the prior, which keeps the candidates close to the compositions already evaluated.
        The architecture, loss, OTC normalization and training schedule are given in SI Section~\ref{SI-si:configs} and SI Table~\ref{SI-tab:configs}.

    \subsection{Surrogate models}
    \label{sec:surrogate_models}

        All DE and GP strategies take the same 17-dimensional composition vector as input, the atomic fractions on the A-site (7 elements) and B-site (10 elements) sublattices, and the vacancy-input variants also take the vacancy count. The output depends on the target: OTC for the scalar targets, all Gibbs energies at once for the Gibbs vector, and one AR energy per composition and vacancy count for vacancy input. MC instead takes the WAE's 8-dimensional latent vector as input and predicts only OTC.
        Training targets are standardized (Z-scores) for all DE strategies and the Gibbs-target GPs, and min--max scaled for scalar GP and MC. The normalization, architectures and training settings are given in SI Section~\ref{SI-si:configs} and SI Table~\ref{SI-tab:configs}.

        \subsubsection{Deep ensemble}%
        \label{sec:de}

            The deep ensemble~\cite{lakshminarayanan2017simple} consists of 16 fully connected networks, each with two hidden layers of 64 units, trained with a Gaussian negative log-likelihood to predict a mean and a variance for each target quantity. For approximate Thompson sampling, one member is drawn at random for each candidate, and its mean prediction is the sample: the OTC score directly for scalar targets, or a set of energies converted to OTC for Gibbs targets. This approximates ensemble sampling~\cite{lu2017ensemble}, which uses a single member for all candidates. Exploration therefore comes from disagreement between the 16 members rather than from their predicted variances.

            Because exploration relies on this disagreement, the members should agree on compositions similar to those already evaluated while continuing to disagree on unfamiliar compositions, rather than converging to nearly the same function everywhere as the database grows. To preserve this spread, each member is regularized toward its own randomly drawn anchor parameters, following approximate Bayesian ensembling~\cite{pearce2020uncertainty}, so that each member behaves approximately like a function drawn from the Bayesian posterior. Approximate Bayesian ensembling trains all members on the same targets, whereas ensemble sampling~\cite{lu2017ensemble} also gives each member its own fixed random perturbation of the targets. Gibbs-vector DE follows ensemble sampling, with a perturbation of the target energies whose standard deviation increases from 0.2 to 0.8~eV with the vacancy count. The scalar and vacancy-input variants use the anchors alone, which are sufficient to keep the spread of the ensemble nearly constant over the 20 cycles (SI Section~\ref{SI-si:configs}).

        \subsubsection{Gaussian process}

            The GP strategies use two model types, a standard single-output GP~\cite{Rasmussen2006} for the scalar and vacancy-input targets and a multi-output GP for the Gibbs vector.
            The single-output GP uses a constant mean, a Mat\'ern-5/2 kernel with a separate length scale for each input dimension and a Gaussian likelihood with a single learned noise variance. Its hyperparameters are fitted by maximizing the exact marginal likelihood, and in the vacancy-input GP the vacancy count receives its own length scale like any other input.
            The multi-output Gibbs-vector GP uses the same kernel type over compositions and couples the 14 nonzero energy outputs (the relative energies at $v=0$ are zero by definition) through a learned covariance between them, using a rank-one intrinsic coregionalization model~\cite{Alvarez2012}.

            Acquisition scores are drawn from the posterior of the underlying function, without the observation noise. The scalar GP draws one OTC value for each candidate. For each candidate, the Gibbs-target GPs draw all energies jointly from the posterior, which keeps the learned correlations between vacancy counts, and convert the draw to OTC from the minima of the two quadratic fits (SI Section~\ref{SI-si:configs}).

        \subsubsection{Monte Carlo dropout}
        \label{sec:mc}

            The MC dropout surrogate~\cite{Gal2016} is the WAE's regression decoder, which predicts OTC from the 8-dimensional latent vector rather than the composition vector.
            The decoder is trained jointly with the rest of the WAE, so surrogate quality is coupled to the WAE representation.
            At inference, dropout remains active, and the empirical mean and variance of 150 stochastic forward passes define a normal distribution from which each candidate's score is drawn.

    \subsection{Prediction targets}
    \label{sec:targets}

        With the scalar targets, the surrogate predicts OTC directly, while with the two Gibbs targets it predicts the relative Gibbs energies, and each sampled set of energies is converted to an OTC score with the quadratic fit. The Gibbs-vector variants give each energy its own output without assuming a common curve, so any relation between the energies has to be learned from the data, either by the multi-output GP or through the shared hidden layers of the DE members. The vacancy-input variants treat the energies as one function of the vacancy count, and the clipped scalar variants apply the capacity ceiling to the sampled score, as described next.

        \subsubsection{Vacancy as an input}
        \label{sec:vacancy_input}

            The vacancy-input variants take the vacancy count as an input next to the composition and predict the AR Gibbs energy $G_{\mathrm{AR}}(x,v)$ at one vacancy count at a time, so the model is evaluated once for each of the eight counts.
            In training, each composition contributes one row for each vacancy count, and all eight rows of a composition are kept on the same side of the training/validation split.

            We chose to model only the AR curve and to derive the FR curve from it, although both curves could also be predicted directly.
            Both curves are computed from the same relaxed structures and phonon calculations, so they differ only through the reactor conditions, with the FR 100~K hotter and at a much lower oxygen partial pressure. Most of the difference between them comes from the change in oxygen chemical potential, and the FR curve is therefore reconstructed as
            \begin{equation}
                \widehat G_{\mathrm{FR}}(x,v)=\widehat G_{\mathrm{AR}}(x,v)-\kappa v,
                \qquad \kappa\approx1.884~\mathrm{eV\ per\ vacancy},
                \label{eq:reactor_relation}
            \end{equation}
            where $-\kappa=\mu_{\mathrm O,FR}-\mu_{\mathrm O,AR}$.
            This relation omits how the vibrational free energy and the configurational entropy term $-TS_{\mathrm{config}}(v)$ change between the two reactor temperatures. Both depend on composition and vacancy count, so the omitted part differs between compositions and grows with the number of vacancies, with a standard deviation across the 1,000 initial compositions that rises from 0.04~eV at one vacancy to 0.09~eV at seven.
            The reconstructed FR energies still reproduce the evaluated ones on these compositions with an RMSE of 0.074~eV and an $R^2$ across compositions of at least 0.998 at each vacancy count (SI Section~\ref{SI-si:prediction}).
            To separate the effect of this approximation on OTC from prediction error, we also apply this relation to the evaluated AR energies and compare the resulting OTC with the OTC from the two evaluated curves (Section~\ref{sec:gibbs_results}).

            During acquisition, each DE member or GP draw gives one AR curve, the same relation gives the matching FR curve, and the quadratic fit turns the pair into one OTC value.
            The reconstruction is used only for scoring, and every selected composition is evaluated with both full reactor curves.

        \subsubsection{Acquisition-score clipping}
        \label{sec:clipping}

            Because OTC scores derived from the Gibbs targets cannot exceed the composition-dependent ceiling $C(x)$, part of their advantage over unclipped scalar prediction could come simply from avoiding unattainable scores above $C(x)$. To isolate this effect, the clipped strategies cap the sampled scalar score $s(x)$ at $C(x)$, so that a scalar prediction cannot rank a candidate above the OTC it would reach by releasing all seven modeled vacancies:
            \begin{equation}
                s_{\mathrm{clip}}(x)=\min\{s(x),C(x)\}.
                \label{eq:score_clipping}
            \end{equation}

            In clipped MC, the cap is computed from a slightly different composition than the one that is later evaluated.
            All clipped strategies compute $C(x)$ after rounding each composition fraction separately to the $1/84$ grid. DE and GP candidates already have integer sublattice counts at this point, so the rounding leaves them unchanged, whereas MC candidates are still continuous, and their rounded compositions have 33 to 38 cations instead of the 36 of the evaluated structure.
            As a result, the cap can differ from the ceiling of the evaluated structure, and clipped MC is reported as run but is only an approximate test of clipping.

    \subsection{Statistical analysis}
    \label{sec:statistics}

        The main measure is each run's mean OTC over new evaluations from cycles 1--20, which reflects how well a search uses its whole evaluation budget and, being averaged over about 200 evaluations, is less sensitive to single noisy evaluations than the other measures. Because a search for oxygen carriers ultimately values its best finds, the mean of the ten best acquired values, the run maximum and threshold counts are reported as secondary measures.
        All are computed from the final databases, counting unique new formulas within each run and excluding the initial dataset, so a formula acquired in several runs counts once in each.
        Run means and upper-tail fractions are divided by each run's number of new evaluations, which is below 200 for some MC runs, while threshold counts are reported per 200 selections.
        For each strategy, the five run values are summarized by their mean and observed range, since a standard deviation is poorly determined from five runs.

        Strategies are compared using one value per AL run for each measure, rather than treating the cycles or acquisitions within a run as independent observations.
        For each measure, 15 primary contrasts fall into four groups (Table~\ref{tab:contrasts} and SI Table~\ref{SI-tab:tests}): clipped against unclipped scalar for each surrogate (3), each Gibbs target against both scalar targets with the same surrogate (8), vacancy input against Gibbs vector for DE and GP (2), and GP against DE for the clipped scalar and Gibbs-vector targets (2).
        The three clipping contrasts are paired by seed because sharing the first-cycle candidate pool keeps their outcomes coupled throughout the search ($r=0.72$ to $0.88$ for run means over cycles 1--20, and $r=0.72$ to $0.90$ over cycles 11--20).
        Runs of other strategies that share a seed label show no such coupling (mean $r=0.01$ over the remaining 42 pairs of strategies), so the other 12 contrasts use unpaired Welch $t$ tests~\cite{welch1947}, which allow the run-to-run variance to differ between strategies.

        Holm adjustment~\cite{holm1979} covers the 15 contrasts within each measure at $\alpha=0.05$, with adjusted $p$ values reported in Table~\ref{tab:contrasts} and unadjusted 95\% intervals in SI Table~\ref{SI-tab:planned}.
        Strategies whose differences are not resolved, that is, with Holm-adjusted $p\geq0.05$, are reported jointly rather than ranked, as recommended for benchmarks with few runs per arm~\cite{Bouthillier2021, Agarwal2021}.
        Because the contrasts were defined after the runs were completed, the analysis is exploratory, and we check how much the conclusions depend on these choices by repeating the clipping contrasts without pairing and by applying Games--Howell comparisons~\cite{games1976} across all 45 pairs of the ten strategies (marked in Table~\ref{tab:contrasts}).
        Tukey comparisons, Westfall--Young max-$T$ permutation adjustment, and exact rank tests, which cannot reach 0.05 at this sample size, are reported in SI Section~\ref{SI-si:stats} and SI Table~\ref{SI-tab:procedures}.

        Two baselines provide reference points outside the hypothesis tests.
        The random baseline comprises the 500 additional compositions whose formulas are not in the initial dataset, with its 95\% range at the AL budget obtained from 20,000 overlapping subsets of 200 drawn without replacement (SI Section~\ref{SI-si:baselines} and SI Table~\ref{SI-tab:random-baseline}).
        The ceiling-only reference runs the active learning protocol without a surrogate, evaluating in each cycle the ten generated candidates with the highest $C(x)$ that pass the screen across five runs completed after the main comparison (SI Section~\ref{SI-si:ceiling}).

    \subsection{Held-out test of the surrogates}
    \label{sec:holdout}

        To compare the targets on common compositions, the DE and GP surrogates were trained once on the initial dataset, with the settings and training split of the first cycle, and used to predict the OTC of compositions they had not seen. The clipped strategies are represented by the scalar prediction capped at $C(x)$, and each model was trained with five seeds. MC is not included, because it predicts only for its own generated latent vectors. The test sets are the 500 random-reference compositions, 58.2\% of which release no oxygen, and the 1,963 compositions acquired by the MC strategies, 65.0\% of which are at the ceiling and only 24 of which were also acquired by a DE or GP strategy. Agreement is measured by the Spearman correlation between predicted and evaluated OTC (SI Section~\ref{SI-si:holdout}).

\section{Results}
    \subsection{Discovery performance across the ten strategies}

        All ten active learning strategies improved substantially on the initial dataset, reaching a mean acquired OTC of 4.50 to 5.32\wt{}, compared with a mean of 0.869\wt{} and a maximum of 4.789\wt{} in the initial dataset (Table~\ref{tab:outcomes}). The 500 additional random compositions show what the same budget of 200 compositions yields without active learning. Like the initial dataset, most of them release no oxygen between the two reactors (58.2\%, against 55.6\% of the initial dataset), none exceeds the initial maximum, and subsets of 200 give a central 95\% range of 0.715--0.992\wt{} for mean OTC (SI Table~\ref{SI-tab:random-baseline}). Every strategy therefore acquires roughly five to six times the mean capacity of an unguided search at the same budget. Much of the gain over random selection appears already in the first cycle (3.86 to 4.77\wt{} on average), once the surrogates avoid the many compositions that release no oxygen, whereas pushing beyond the initial dataset over the full 20 cycles depends strongly on the strategy.

\begin{figure}[p]
            \centering
            \includegraphics[width=\textwidth]{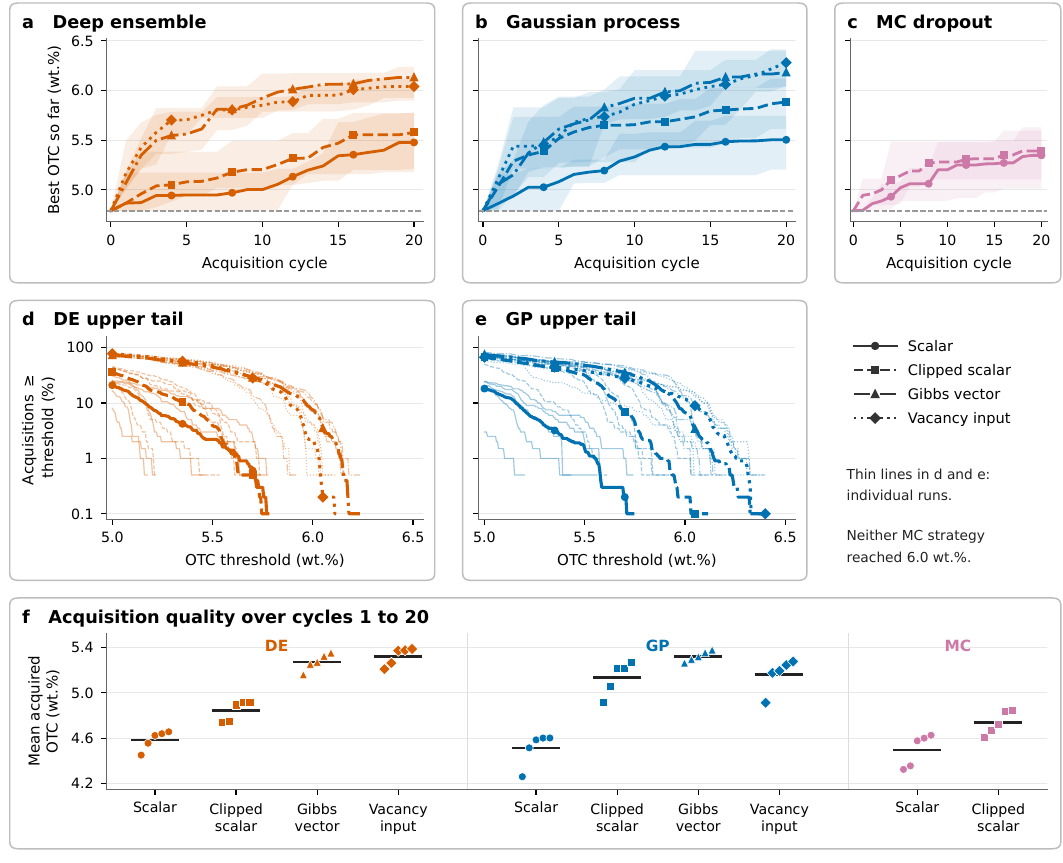}
            \caption{\textbf{Discovery performance.}
            \textbf{a--c,} Best OTC found through each cycle, including the initial set, for the DE, GP and MC families. Lines are five-run means, and shading spans the run minimum and maximum. The dashed line is the common initial maximum.
            \textbf{d,e,} Fraction of new evaluations at or above each OTC threshold, for the DE strategies in d and the GP strategies in e. These performance profiles~\cite{Agarwal2021} show the whole upper tail of the outcomes rather than a single summary. Thin lines are individual runs and thick lines five-run means. Thresholds reached by no acquisition leave gaps on the logarithmic axis. Neither MC strategy acquired OTC $\geq6\wt{}$, and SI Fig.~\ref{SI-fig:s03} shows their tails on a linear scale.
            \textbf{f,} Mean acquired OTC over cycles 1--20. Each symbol is one run and each black line is the five-run mean. The ceiling-only reference lies below the axis, at 1.79\wt{} (run range 0.14--2.62\wt{}, Fig.~\ref{fig:surrogate_value}a).}
            \label{fig:discovery}
        \end{figure}

        Across the 20 cycles, changing what the surrogate predicts affects mean acquired OTC far more than changing the surrogate itself. Every one of the 20 runs with a Gibbs target acquired a higher mean OTC (4.91 to 5.39\wt{}) than every one of the 15 runs with an unclipped scalar target (4.26 to 4.66\wt{}). Predicting unclipped OTC directly, DE, GP and MC reach five-run means within 0.09 percentage points of one another, 4.50 to 4.58\wt{}. Predicting Gibbs energies instead raises the five-run mean by 0.65 to 0.81 percentage points with the same surrogate, putting the four DE and GP Gibbs-target strategies in a leading group at 5.16 to 5.32\wt{}, with mean run maxima of 6.04 to 6.28\wt{} against 5.35 to 5.50\wt{} for the unclipped scalar strategies (Fig.~\ref{fig:discovery}). Neither the primary contrasts (Table~\ref{tab:contrasts}) nor the Games--Howell comparisons (SI Table~\ref{SI-tab:stats-all}) separate the four Gibbs-target strategies on either measure, and on mean acquired OTC clipped GP (5.13\wt{}, Section~\ref{sec:clipping_results}) is not separated from them either.

        The four Gibbs-target strategies also account for 279 of the 281 acquisitions at or above $6\wt{}$ (Fig.~\ref{fig:discovery}d,e), a reference level between the stoichiometric capacities of the single-metal oxygen carriers Fe$_2$O$_3$/Fe$_3$O$_4$ and Mn$_3$O$_4$/MnO quoted in the Introduction. Counted per run, the number of acquisitions at or above $6\wt{}$ has a median of 34 for vacancy-input GP, 15 for Gibbs-vector DE, 10 for Gibbs-vector GP and 3 for vacancy-input DE, against 0 for clipped GP and every other scalar strategy (Table~\ref{tab:outcomes}). These counts vary strongly between runs, from 0 to 24 for Gibbs-vector DE and 4 to 39 for vacancy-input GP. The order of the four also reverses at $5\wt{}$, where vacancy-input DE leads with a mean of 155.4 acquisitions per run and vacancy-input GP is last with a mean of 133.6. Which of them leads therefore depends on whether the goal is many good candidates or a few exceptional ones. Individual outcomes and trajectories are given in SI Table~\ref{SI-tab:runs} and SI Figs.~\ref{SI-fig:s01}--\ref{SI-fig:s03}.

        \begin{table}[htbp]\centering\small
            \caption{Outcomes over $n=5$ runs per strategy. OTC entries are the mean of the five run summaries followed by the observed run range, in wt.\%.
            The maximum column summarizes run maxima rather than reporting the largest pooled observation. Yield is the median number of new evaluations at or above $6\wt{}$ per run of 200 selections, with the run range in parentheses. The median is reported because some strategies have runs with very different counts.}
            \label{tab:outcomes}
            \begin{tabular}{lccr}\toprule
                Strategy & Mean acquired OTC & Run maximum OTC & Yield $\geq6\wt{}$\\
                & mean (range) & mean (range) & median (range)\\\midrule
                DE scalar & 4.58 (4.45--4.66) & 5.48 (5.18--5.78) & 0 (0--0) \\
                DE clipped & 4.84 (4.74--4.92) & 5.58 (5.21--5.77) & 0 (0--0) \\
                DE Gibbs vector & 5.27 (5.16--5.35) & 6.13 (5.92--6.24) & 15 (0--24) \\
                DE vacancy input & 5.32 (5.21--5.39) & 6.04 (5.90--6.12) & 3 (0--8) \\
                GP scalar & 4.51 (4.26--4.60) & 5.50 (5.20--5.74) & 0 (0--0) \\
                GP clipped & 5.13 (4.92--5.27) & 5.88 (5.61--6.12) & 0 (0--2) \\
                GP Gibbs vector & 5.32 (5.26--5.38) & 6.18 (6.03--6.40) & 10 (3--27) \\
                GP vacancy input & 5.16 (4.91--5.28) & 6.28 (6.05--6.41) & 34 (4--39) \\
                MC scalar & 4.50 (4.32--4.63) & 5.35 (5.01--5.63) & 0 (0--0) \\
                MC clipped & 4.74 (4.61--4.85) & 5.39 (5.11--5.60) & 0 (0--0) \\
                \bottomrule\end{tabular}
        \end{table}

        \begin{table}[htbp]\centering\small
            \caption{Primary contrasts between strategies ($n=5$ runs per strategy). Each row compares the strategy in the first column with the one in the second. Differences are the first minus the second in OTC percentage points, and $p$ is two-sided, Holm-adjusted over the 15 contrasts in each column. Values below 0.05 are in bold and mark differences resolved in the primary analysis. Clipping contrasts are paired by seed, and the others use Welch $t$ tests (SI Table~\ref{SI-tab:tests}). Intervals and sensitivity analyses are given in SI Tables~\ref{SI-tab:planned}, \ref{SI-tab:stats-all} and \ref{SI-tab:procedures}.}
            \label{tab:contrasts}
            \setlength{\tabcolsep}{4pt}
            \begin{tabular}{llrrrrrr}\toprule
                & & \multicolumn{2}{c}{Mean acquired OTC} & \multicolumn{2}{c}{Run maximum OTC} & \multicolumn{2}{c}{Top-ten mean OTC}\\
                \cmidrule(lr){3-4}\cmidrule(lr){5-6}\cmidrule(lr){7-8}
                Strategy & Compared with & Diff. & $p$ & Diff. & $p$ & Diff. & $p$\\\midrule
                \multicolumn{8}{l}{\emph{Clipped against unclipped scalar, paired by seed}}\\
                DE clipped & DE scalar & $+0.26$ & \textbf{0.0017} & $+0.10$ & 0.96 & $+0.11$ & 0.78 \\
                GP clipped & GP scalar & $+0.62$ & \textbf{0.0017} & $+0.38$ & \textbf{0.039}$^{\dagger}$ & $+0.40$ & 0.096 \\
                MC clipped & MC scalar & $+0.24$ & \textbf{0.012}$^{\dagger}$ & $+0.04$ & 1.00 & $+0.07$ & 1.00 \\
                \multicolumn{8}{l}{\emph{Gibbs target against scalar}}\\
                DE Gibbs vector & DE scalar & $+0.69$ & \textbf{\textless 0.001} & $+0.65$ & \textbf{0.047}$^{\dagger}$ & $+0.71$ & \textbf{0.023} \\
                DE vacancy input & DE scalar & $+0.74$ & \textbf{\textless 0.001} & $+0.56$ & 0.074 & $+0.62$ & \textbf{0.042}$^{\dagger}$ \\
                DE Gibbs vector & DE clipped & $+0.43$ & \textbf{\textless 0.001} & $+0.55$ & \textbf{0.036} & $+0.59$ & \textbf{0.013} \\
                DE vacancy input & DE clipped & $+0.48$ & \textbf{\textless 0.001} & $+0.46$ & 0.065 & $+0.51$ & \textbf{0.031} \\
                GP Gibbs vector & GP scalar & $+0.81$ & \textbf{0.0011} & $+0.68$ & \textbf{0.0061} & $+0.73$ & \textbf{0.0082} \\
                GP vacancy input & GP scalar & $+0.65$ & \textbf{0.0011} & $+0.78$ & \textbf{0.0028} & $+0.80$ & \textbf{0.0038} \\
                GP Gibbs vector & GP clipped & $+0.19$ & 0.21 & $+0.30$ & 0.15 & $+0.33$ & \textbf{0.042}$^{\dagger}$ \\
                GP vacancy input & GP clipped & $+0.03$ & 0.78 & $+0.40$ & 0.055 & $+0.40$ & \textbf{0.026} \\
                \multicolumn{8}{l}{\emph{Vacancy input against Gibbs vector}}\\
                DE vacancy input & DE Gibbs vector & $+0.05$ & 0.72 & $-0.09$ & 0.96 & $-0.09$ & 0.91 \\
                GP vacancy input & GP Gibbs vector & $-0.16$ & 0.26 & $+0.10$ & 0.96 & $+0.07$ & 1.00 \\
                \multicolumn{8}{l}{\emph{GP against DE}}\\
                GP clipped & DE clipped & $+0.29$ & \textbf{0.044}$^{\dagger}$ & $+0.31$ & 0.28 & $+0.28$ & 0.24 \\
                GP Gibbs vector & DE Gibbs vector & $+0.05$ & 0.72 & $+0.05$ & 1.00 & $+0.03$ & 1.00 \\
                \bottomrule\end{tabular}

                \smallskip
                \parbox{\textwidth}{\footnotesize $^{\dagger}$Not resolved by Games--Howell comparisons over all 45 pairs of strategies.}
        \end{table}

    \subsection{Effect of acquisition-score clipping}
    \label{sec:clipping_results}

        Clipping tests how much of the gap between the scalar and Gibbs targets is closed simply by removing acquisition scores above the capacity ceiling. It improves the mean OTC of scalar-model acquisitions for all three surrogates, with the largest gain for GP (Table~\ref{tab:contrasts}). Compared with the unclipped strategy at the same seed, mean acquired OTC rises by 0.26 percentage points for DE, 0.62 for GP and 0.24 for MC. The MC gain is resolved only when runs are paired by seed. Clipping mainly raises the body of the distribution rather than its top: no clipping gain is resolved on the mean of the ten best acquisitions for any surrogate, and on the run maximum only the GP gain is resolved, and only in the primary analysis. At a $5\wt{}$ threshold clipped GP delivers as many acquisitions as the Gibbs-target strategies, 134.6 per run on average against 133.6 to 155.4, but unlike them it rarely reaches $6\wt{}$ (Table~\ref{tab:outcomes}).

        The larger gain for GP is consistent with how often its unclipped scores exceed the structure's ceiling, in 96.5\% of its selections compared with 51.0\% for DE, since these unattainable scores are what clipping removes (Fig.~\ref{fig:mechanism}a). Clipped DE and GP never exceed the ceiling, and neither does any Gibbs-target strategy. For 52.4\% of clipped-GP selections, the clipped score equals the composition-dependent ceiling $C(x)$, so these candidates are ranked by composition mass, lightest first (Section~\ref{sec:compositions}).

        Clipped MC uses a separately rounded threshold mass, and 42.6\% of its selected scores still exceed the ceiling of the saved structure, against 86.2\% for unclipped MC. Its clipping therefore removes only part of the excess and tests a weaker form of the same idea (SI Section~\ref{SI-si:mc_audit} and SI Fig.~\ref{SI-fig:s04}).

    \subsection{Performance of Gibbs-energy representations}
    \label{sec:gibbs_results}
        Predicting the Gibbs energies improves on scalar clipping at the top of the distribution for both surrogates, and across all acquisitions for DE (Table~\ref{tab:contrasts}). On the mean of the ten best acquisitions, the differences from clipped scalar are 0.59 and 0.51 percentage points for DE and 0.33 and 0.40 points for GP, and three of the four are also resolved under Games--Howell. On mean acquired OTC they are 0.43 and 0.48 points for DE, but only 0.19 and 0.03 points for GP, and neither GP difference is resolved.

        Between the two Gibbs representations themselves, no difference is resolved on any of the three measures for either surrogate, with mean acquired OTC differing by only 0.05 percentage points for DE and 0.16 points for GP (Table~\ref{tab:contrasts} and SI Table~\ref{SI-tab:planned}). The linear reactor relation used by the vacancy-input variants (equation~\eqref{eq:reactor_relation}) also introduces too little error to affect this comparison, changing OTC on the initial dataset by only 0.016 percentage points on average, or at most 0.61 points for a single composition (SI Fig.~\ref{SI-fig:s06}).

    \subsection{Why the leading strategies win: mass and saturation}
    \label{sec:compositions}
\begin{figure}[p]
            \centering
            \includegraphics[width=\textwidth]{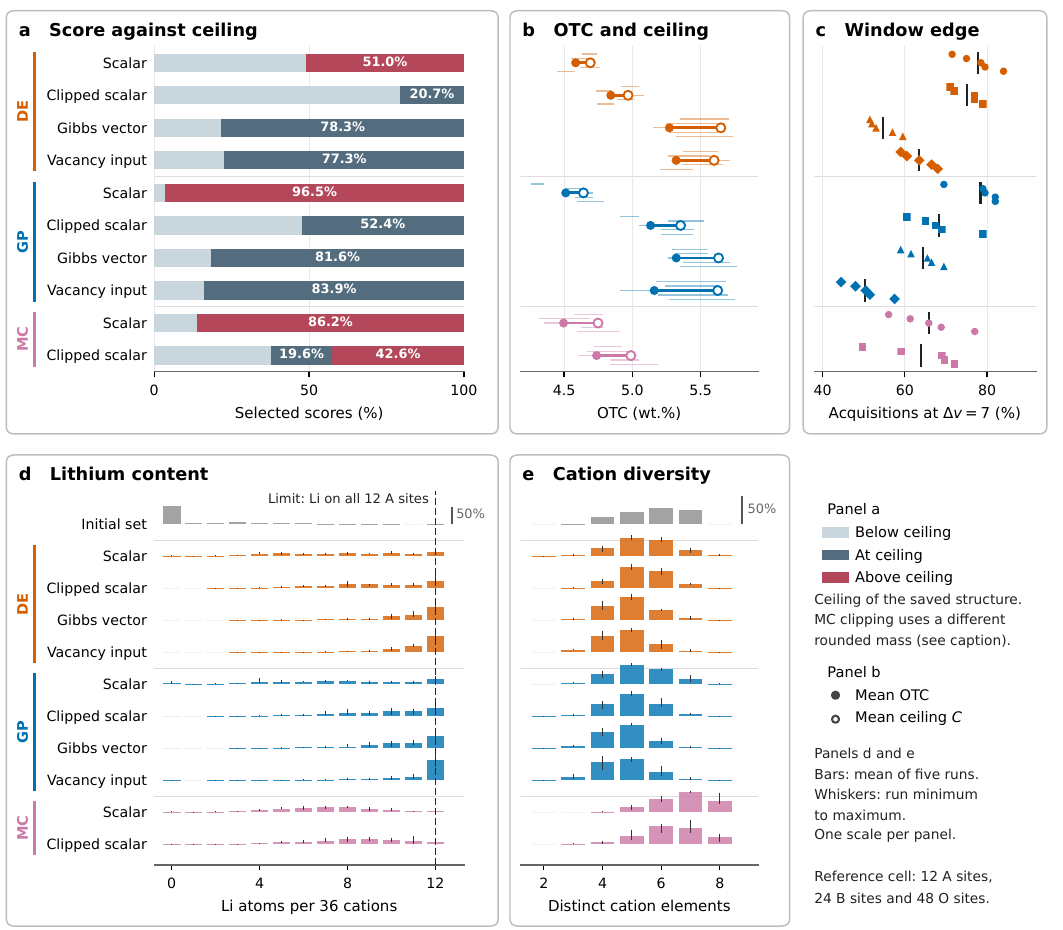}
            \caption{\textbf{Acquisition scores, capacity ceiling and selected chemistry.}
            \textbf{a,} Share of selected acquisition scores below, at and above the ceiling $C(x)$ of the evaluated structure, pooled over the five runs of each strategy. Clipped MC computes its cap from a separately rounded composition (Section 2.8.2), so some of its selected scores still lie above this ceiling.
            \textbf{b,} Mean evaluated OTC (filled circles) and mean ceiling (open circles). The bold line joining them is the five-run mean, and the thin lines show the same span for each run, so the length of a line is the mean shortfall below the ceiling. The ceiling-only reference lies outside the axis (Fig.~\ref{fig:surrogate_value}a).
            \textbf{c,} Share of each run's acquisitions that release all seven modeled vacancies ($\Delta v=7$), with black marks at the five-run means.
            \textbf{d,e,} Distributions of Li count and number of distinct cation elements. Bars are mean run fractions, whiskers span the five runs, and gray bars show the initial dataset.}
            \label{fig:mechanism}
        \end{figure}

        The differences in capacity between strategies come largely from the masses of the compositions they select, because most selected compositions release all seven vacancies of the modeled interval, so that their OTC equals the ceiling $C(x)$ (Fig.~\ref{fig:mechanism}c). This holds for 66.2\% of the 9,982 new evaluation records and for 89.3\% of the 281 records with OTC $\geq6\wt{}$. The ranking of strategies is therefore largely a ranking of how quickly each search moves to light compositions that still release their oxygen. Across the 50 runs, a run's mean OTC closely follows the mean ceiling of the compositions it selects (Pearson $r=0.974$, Figs.~\ref{fig:mechanism}b and \ref{fig:surrogate_value}a, SI Table~\ref{SI-tab:blocks}), whereas runs that reach a larger fraction of their ceiling tend to have lower mean OTC ($r=-0.465$). Low mass alone would not have identified these compositions in the initial dataset, however, where ceiling and OTC are essentially uncorrelated (Pearson $r=-0.010$) and the ten highest-ceiling compositions have a mean OTC of only 0.58\wt{}, against 4.63\wt{} for the ten highest-OTC compositions.

    \subsection{What the surrogate adds beyond selecting for low mass}
    \label{sec:surrogate_value}
\begin{figure}[htbp]
            \centering
            \includegraphics[width=\textwidth]{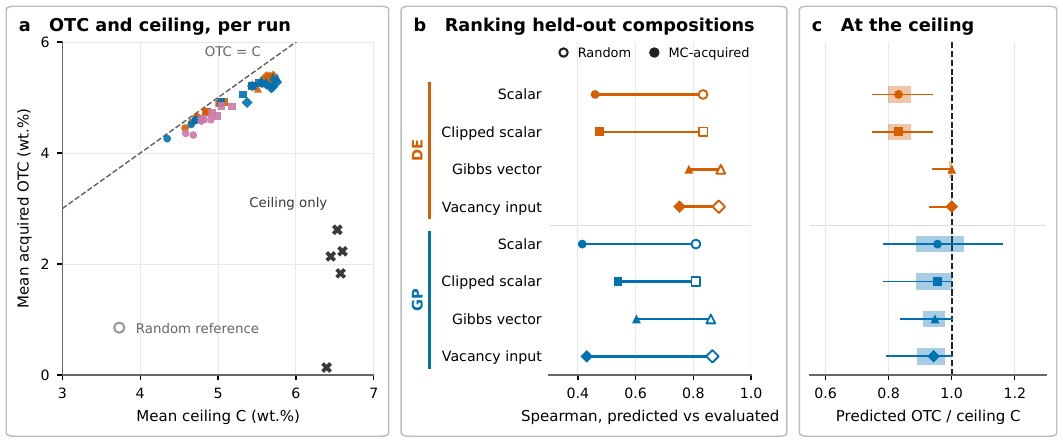}
            \caption{\textbf{Ranking by mass alone and held-out accuracy of the surrogates.}
            \textbf{a,} Mean acquired OTC against mean ceiling $C$, one point per run, for the ten strategies (markers as in Fig.~\ref{fig:discovery}), the ceiling-only reference (crosses) and the random reference (open circle). Dashed line: OTC $=C$.
            \textbf{b,} Spearman correlation between predicted and evaluated OTC in the held-out test, for the random-reference (open) and MC-acquired (filled) compositions. Means over five training seeds, which differ by at most 0.04. Clipped scalar is the scalar prediction capped at $C$.
            \textbf{c,} Predicted OTC as a fraction of the ceiling $C$ for the 1,275 MC-acquired compositions at the ceiling, pooled over the five seeds. The symbol marks the median, the box spans the middle 50\% of the values and the line the middle 90\%.}
            \label{fig:surrogate_value}
        \end{figure}

        The ceiling-only reference tests what happens when a search ranks candidates by mass alone. It reaches the highest mean ceiling of any search, 6.51\wt{}, but a mean acquired OTC of only 1.79\wt{}, less than half that of any strategy, because only 0.3\% of its acquisitions release all seven vacancies (Fig.~\ref{fig:surrogate_value}a and SI Section~\ref{SI-si:ceiling}). Ranking by mass can still reach the best compositions by chance. Two of its five runs found compositions at 6.40 and 6.41\wt{}, as high as any found by the ten strategies, so its run maxima are not resolved from theirs. These are only 4 of its 1,000 selections at or above $6\wt{}$, however, and the other three runs stay below $5.5\wt{}$. The surrogates therefore contribute by reliably identifying which light compositions also release their oxygen. Prediction errors logged during acquisition cannot rank how well the surrogates do this, because each strategy selects different compositions (SI Table~\ref{SI-tab:prediction}), so we compare the surrogates on common held-out sets instead (Section~\ref{sec:holdout}). Trained only on the initial dataset, all surrogates rank the held-out random-reference compositions well, with Spearman correlations of 0.81 to 0.90 (Fig.~\ref{fig:surrogate_value}b). On the compositions acquired by the MC strategies, which resemble those the searches select, the correlations of the scalar-target surrogates fall to 0.42 to 0.54, as these surrogates are least accurate at the two bounds of OTC, for compositions that release no oxygen and for those at the ceiling (Fig.~\ref{fig:surrogate_value}c and SI Table~\ref{SI-tab:holdout}). The Gibbs-target DEs reach correlations of 0.75 and 0.78 on this set, whereas the Gibbs-target GPs reach only 0.60 and 0.43. The difference arises mainly among compositions at the ceiling, where averaging OTC over the GP posterior samples places the predictions below $C(x)$ (SI Section~\ref{SI-si:holdout}), so the held-out test does not fully account for search performance.

    \subsection{Favored compositions and their chemical feasibility}
    \label{sec:feasibility}
        Once a composition releases all seven modeled vacancies, its OTC can increase only through a lower mass, by replacing heavier cations with lighter ones while it still releases all seven. Repeated over many cycles, this favors the lightest cations, Li and Al, and the search therefore moves toward simpler mixtures: mean cation mixing entropy $S_{\mathrm{mix}}$ falls from $2.67R$ in the initial dataset to $1.97R$ among acquisitions at or above $6\wt{}$, and 93\% of these contain four or five cation species (Fig.~\ref{fig:mechanism}e). Most of them still exceed $1.5R$, the threshold commonly used to define high-entropy compositions (SI Fig.~\ref{SI-fig:s05}), but that criterion sets a low bar here because it sums over three cation sites per formula unit, where an equimolar two-element B sublattice alone gives $2\ln 2\approx1.39R$.

        Packing Li onto the A sublattice and Al onto the B sublattice is most pronounced in the leading strategies. For DE, the mean Li count per structure rises from 7.6 of the 12 A sites with scalar prediction to 9.2 with clipping and 10.6 with Gibbs-vector prediction (Fig.~\ref{fig:mechanism}d). The thirty highest-OTC acquisitions are all Li--Al--Mn compositions with three to five cation species, placing Li on 11 or 12 A sites and 13 to 16 Al on the B sublattice. That leaves only 8 to 11 redox-active cations, mostly Mn, which must average an oxidation state of $+4.0$ to $+4.5$ in the oxidized cell and take up 14 electrons to release seven oxygen atoms. In the extreme case, the lightest composition the screen allows, Li$_{12}$Al$_{21}$Mn$_{3}$O$_{48}$ ($C=7.08\wt{}$, against a maximum of $5.73\wt{}$ in the initial dataset, SI Section~\ref{SI-si:baselines}), requires Mn(VII).

        Because the acquisition screen allows oxidation states up to Mn(VII), Cr(VI) and Fe(VI) (SI Table~\ref{SI-tab:screen_states}), we checked retrospectively how many acquisitions can still balance the 48 oxygen ions when oxidation states are capped at the values usual in spinels, Mn(IV), Cr/Fe/Co/Ni(III) and Cu(II) (SI Section~\ref{SI-si:screening}). Under this check, 70.0\% of the 200 highest-OTC acquisitions cannot balance charge, compared with 8.5\% of the 200 highest-OTC initial compositions. Across all acquisitions, the share is 28.6\%, against 2.5\% for the initial dataset. Even when Cu(III), Co/Ni/Fe(IV) and Cr(VI) are also allowed, 12.5\% of the top 200 acquisitions still cannot balance charge, as they would need Mn or Fe beyond the IV state. Among the acquisitions that pass the stricter check, all four Gibbs-target strategies still exceed every unclipped scalar strategy in mean OTC. Vacancy-input GP changes the most. Its mean OTC falls below that of clipped GP, and its mean number of acquisitions at or above $6\wt{}$ drops from 23.4 to 6.2 per run, just behind Gibbs-vector DE at 6.6 (SI Table~\ref{SI-tab:charge}).

\section{Discussion}

    \subsection{Predicting Gibbs energies instead of OTC}
    \label{sec:practitioner}

        We recommend predicting the intermediate Gibbs energies, because both DE and GP perform well with either Gibbs representation, whereas the performance of clipped scalar models depends strongly on the surrogate. Reducing each pair of energy curves to a single OTC value discards the shape of the curves, and this loss of information is most severe at the two hard bounds of OTC, at zero for 55.6\% of the initial dataset and at the ceiling for 66.2\% of the acquisitions. At zero OTC, a scalar surrogate receives the same value whether a composition's FR curve rises steeply with vacancy count or lies nearly flat, just above the threshold for releasing oxygen. Likewise, for compositions at the ceiling, a scalar surrogate receives only the mass-fixed limit $C(x)$, without distinguishing a curve whose FR minimum barely reaches seven vacancies from one that drops strongly across the interval, or an AR minimum that sits safely at zero from one close to opening vacancies in air. The scalar surrogates fit these piled-up values with a Gaussian likelihood, and the held-out test shows that they are least accurate exactly at the two bounds. The Gibbs energies have no such bounds, and although each evaluation is noisy because it uses a single SQS and vacancy arrangement, the underlying energies can be expected to vary smoothly with composition, so they preserve these differences as training signals.

        Predicting the energies also adds little computational overhead, and because DE and GP perform similarly with the Gibbs targets, the choice between them is mainly practical. Even for the slowest Gibbs-target strategy, data loading and surrogate fitting take under 3\% of the estimated GPU time per cycle, against about 95\% for candidate evaluation (SI Table~\ref{SI-tab:timings}). At the dataset sizes of this study, the Gibbs-target GPs trained faster than the corresponding DEs and required fewer architectural and training choices, whereas on much larger datasets the linear scaling of a DE or the factored covariance of the Gibbs-vector GP is preferable to the cubic scaling of an exact GP over composition--vacancy rows~\cite{Rasmussen2006, Burt2020}.

    \subsection{Why clipping helps, and where it stops}
    \label{sec:why_clipping}

        For the scalar targets, clipping removes sampled scores above a composition's capacity ceiling, and its larger gain for GP is consistent with how much more often the unclipped GP scores exceeded the ceiling. Part of this excess is already present in the predictive mean, which lies above the ceiling for 43.5\% of GP selections and 27.6\% of DE selections, and sampling adds the rest. Because only the ten highest of 25,000 sampled scores are selected in each cycle, the selected GP scores lay a median 0.69 percentage points above the predictive mean, against 0.14 for DE. The GP uncertainty was about four times as wide, and a Gaussian draw also reaches further into the tail than a draw from 16 ensemble members, with 10.2\% of GP selections more than three standard deviations above the mean, against 0.6\% for DE. Candidates whose scores are clipped, however, are ranked by their ceilings and therefore by molar mass alone. That ordering is accurate only when the candidates actually release all seven vacancies, which is why ranking by mass alone rarely finds high-capacity compositions and helps explain why clipping does not match the Gibbs targets among the highest-capacity discoveries.

    \subsection{How the screen shapes the selected compositions}
    \label{sec:screen}

        Because the search favors low composition mass, it increasingly selects Li- and Al-rich compositions. With so few redox-active cations, whether these compositions pass the screen depends on the oxidation states it allows. This illustrates how design assumptions can shape active learning outcomes~\cite{Nair2026}, and why a screen should be checked against the compositions a search ends up selecting, not only against the initial dataset.

        Because the screen and the seven-vacancy limit shape the selected compositions so strongly, the highest evaluated capacities, up to 6.4\wt{}, are best read as outcomes of the chosen objective rather than as recommended carriers. They approach the stoichiometric capacity of the Mn$_3$O$_4$/MnO couple quoted in the Introduction, but they are bounded by the seven-vacancy limit, and compositions that remain stable beyond it could release more oxygen, while those that transform earlier would release less. The highest-capacity candidates also owe part of their capacity to a high Li content. The modest gain from a lower composition mass is unlikely to justify the raw-material cost of Li or the risk of Li loss at the reactor temperatures, since the Li spinel lithium ferrite loses lithium oxide above about 1,250~K, largely as a consequence of oxygen loss~\cite{Ridgley1970}. Candidates that pass the stricter charge check would still need to be assessed for phase stability and reversible oxygen release across the modeled vacancy range under the reactor conditions.

        OTC is a standard measure for comparing oxygen carriers, but it leaves out properties that matter in practice. These include stabilization through configurational entropy, which is the main motivation for high-entropy carriers, the rates of reduction and oxidation, resistance to attrition and agglomeration, raw-material cost, availability and supply-chain criticality, environmental impact, and the stability of the oxidized and reduced states under reactor conditions. The comparison measures how efficiently each strategy reaches the chosen objective, while a search for practical carriers would likely benefit from an objective that also accounts for these properties. The OTC is also evaluated at a single temperature and oxygen partial pressure per reactor, whereas in practice the temperatures are limited by the carrier itself and the gas composition changes from inlet to outlet.

    \subsection{Scope of the comparison}
    \label{sec:limitations}

        The advantage of predicting intermediate energies appears across both surrogate families and both Gibbs representations, supporting this approach beyond a single model implementation. At the same time, all strategies were tested with Thompson sampling, a shared 1,000-composition initial dataset and five runs per strategy, so the comparison measures variation from a common starting point under one acquisition rule, and whether the size of the effect of the prediction target changes under other acquisition functions or for objectives not dominated by the seven-vacancy limit remains to be tested. The per-candidate Thompson draws (Section 2.1) may also matter here. With 25,000 candidates, the selected scores come from the far tail of the predictive distributions. This inflates the scores most when OTC is predicted directly (Section 4.2), so joint posterior samples or another acquisition function could change how much clipping helps. Similarly, because every strategy is evaluated with the same CHGNet-based pipeline, the comparison between strategies is internally consistent, whereas confirming the capacities of specific candidate compositions would require DFT calculations and experimental testing.

    \subsection{Outlook: intermediate targets in active learning}
    \label{sec:outlook}

        The same choice of prediction target arises across computational materials discovery whenever a target property is computed from smoother intermediate quantities through a known physical mapping, as in Bayesian optimization of composite functions~\cite{Astudillo2019, Cardoso_Coelho_2025}. Common examples include catalytic activity or overpotential derived from elementary reaction and adsorption energies~\cite{Norskov2009, Tran_2018}, phase stability determined from the formation energies of competing phases~\cite{MeBaSc2023, Novick_2024}, intercalation voltages computed from formation energies across Li compositions~\cite{Urban_2016}, and the stable charge state of a defect, set by the lowest formation energy across charge states~\cite{Freysoldt_2014}. Some active learning studies already predict such intermediate quantities, for example adsorption energies for catalyst screening~\cite{Tran_2018} or formation energies whose uncertainty is propagated through the convex hull~\cite{Novick_2024}, whereas the comparison here separates the effect of this choice from that of the surrogate model. Because any mapping from a vector of intermediate quantities to a single scalar is many-to-one, training a surrogate directly on the final scalar discards how the individual quantities vary with composition, and when the mapping also includes minima, thresholds or bounds, it flattens inactive or saturated regions into identical values. Predicting the intermediate quantities instead lets the surrogate learn smooth composition--energy relationships while applying the known physical mapping during acquisition.
        The gain is not guaranteed, however. When the mapping takes small differences between large intermediate quantities, prediction errors can be amplified. For example, compositional models that predict formation energies well can still predict compound stability poorly \cite{Bartel2020}. Intermediate targets should help most when the mapping does not amplify small errors in the predicted quantities.

        This approach is practical whenever the evaluation pipeline already computes the intermediate quantities on the way to the target property, as it requires no extra evaluations and adds little surrogate training cost compared with atomistic or first-principles calculations. Our comparison between the Gibbs-vector and vacancy-input strategies also suggests that the surrogate need not predict every condition as a separate output. Where a known thermodynamic relation links conditions, such as the oxygen chemical potential shift between reactors here or temperature and pressure terms in a free-energy model, building that relation into the prediction target lets a simpler model achieve comparable discovery gains.

\section{Conclusion}

    Across ten active learning strategies for discovering spinel oxygen carriers, what the surrogate model was trained to predict mattered far more than which surrogate model was used. The reason is that most evaluated compositions have a capacity of zero or of their own ceiling, so a scalar capacity target gives the same value to compositions whose Gibbs energy curves can differ widely. Predicting the Gibbs energies lets the surrogates learn from these differences, and the deep ensembles and Gaussian processes trained this way made 279 of the 281 acquisitions at or above 6 wt.\%, although the count varied strongly between runs. Clipping the scalar scores at the ceiling removes unattainable scores but then ranks the clipped candidates by molar mass alone, so it narrows the gap in mean capacity without matching the Gibbs targets among the best discoveries.

    Analyzing the selected compositions shows both what the surrogates learn and how the objective shapes the search. Ranking candidates by molar mass alone yielded less than half the mean capacity of any active learning strategy, confirming that the surrogates add value by identifying which light compositions actually release their oxygen. At the same time, because most selected carriers saturate the modeled seven-vacancy interval, maximizing capacity per unit mass drives every strategy toward Li- and Al-rich spinels at the limits of the oxidation-state screen, showing how an active search will exploit permissive screening rules when the optimum lies at the boundary of the allowed space. For active learning campaigns more broadly, when a target property is a composite function of intermediate physical quantities, predicting those intermediate quantities can matter far more for discovery performance than the choice of surrogate model.

\section*{Author contributions}
    A.J.O. developed the deep ensemble surrogate, the quadratic-fit target extraction, and performance improvements to the active learning pipeline, designed the strategy comparison, performed all calculations, and wrote the manuscript. I.S. and R.S.Y. contributed to discussions of the methodology and results and reviewed the manuscript. T.M. and A.H. supervised the work and reviewed and edited the manuscript. All authors discussed the results and commented on the paper.

\section*{Conflicts of interest}
    There are no conflicts to declare.
    
\section*{Data availability}
    All data generated and analyzed in this study are available in the Zenodo repository at \url{https://doi.org/10.5281/zenodo.21283967}. The record contains the training databases (ASE format) of all 50 runs, the five ceiling-only runs, the initial dataset and the random reference, together with the relaxed structures and phonon data of every evaluated composition. It also contains the configuration, log and per-cycle candidate lists of each run, the results of the held-out test, the derived tables and the figures, and the container image of the software environment of the runs.

    The code of this study is available at \url{https://github.com/alfredjo00/spinel-al} under the MIT license. It contains the active learning loop with all surrogates and prediction targets, the code that screens compositions and evaluates them (SQS generation, CHGNet relaxation, Phonopy calculations, and the Gibbs energy and OTC extraction), the held-out test, and the analysis and plotting scripts that reproduce all figures, tables and statistics of the article and the Supplementary Information from the Zenodo data. The first commit of the repository is the code as used for the runs.

\section*{Acknowledgments}
    This work was funded by the Swedish Research Council (2020-03487) and the Chalmers Area of Advance Energy. The computations were enabled by resources provided by the National Academic Infrastructure for Supercomputing in Sweden (NAISS) at NSC and C3SE, partially funded by the Swedish Research Council through grant agreement no. 2022-06725, under projects NAISS 2025/5-313, NAISS 2025/5-530, NAISS 2026/3-535, NAISS 2026/3-498 and NAISS 2026/3-291, and by the Berzelius resource provided by the Knut and Alice Wallenberg Foundation at the National Supercomputer Centre. The funders played no role in study design, data collection, analysis and interpretation of data, or the writing of this manuscript.

\begingroup\small
\setlength{\bibsep}{3pt}
\bibliographystyle{naturemag}
\bibliography{manuscript}
\endgroup
\clearpage
\phantomsection\hypertarget{si-start}{}
\includepdf[pages=-]{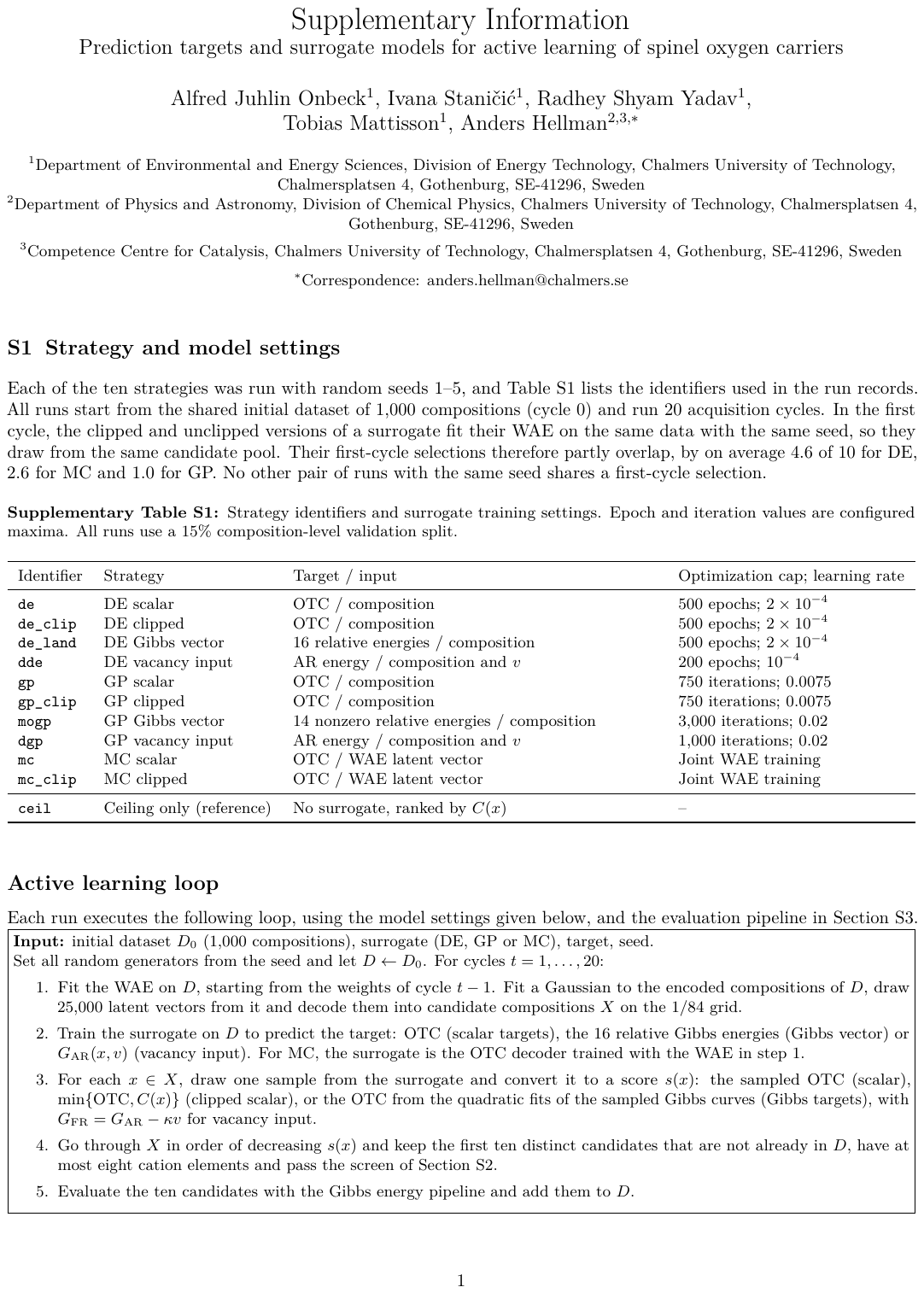}
\end{document}